\documentclass[preprint,12pt]{elsarticle}
\usepackage{graphics}
\usepackage{rotating}
\usepackage{mathrsfs}
\usepackage{amsmath}
\usepackage{lscape}
\usepackage[usenames,dvipsnames]{color}
\usepackage{tikz,ifthen,calc}
\usetikzlibrary{snakes,shapes}
\usetikzlibrary{patterns}
\usetikzlibrary{plotmarks}
\usetikzlibrary{calc}
\usepackage{pgfplots}
\newcommand{\nn}{\nonumber\\}
\newcommand{\be}{\begin{equation}}
\newcommand{\ee}{\end{equation}}
\newcommand{\bea}{\begin{eqnarray}}
\newcommand{\eea}{\end{eqnarray}}

\journal{}
\begin{document}
\begin{frontmatter}
\title{Multistep shell model description of spin-aligned neutron-proton pair coupling}

\author{Z.X. Xu}
\author{C. Qi\corref{email}}
\author{J. Blomqvist}
\author{R. J. Liotta}
\author{R. Wyss}
\address{Royal Institute of Technology (KTH), Alba Nova University Center,
SE-10691 Stockholm, Sweden}
\cortext[email]{Corresponding author.\\
\textit{E-mail address:} chongq@kth.se (Chong Qi)}

\begin{abstract}
The recently proposed spin-aligned neutron-proton pair coupling scheme is studied within a non-orthogonal basis in term of the multistep shell model. This allows us to identify simultaneously the roles played by other configurations such as the normal pairing term.
The model is applied to four-, six- and eight-hole $N=Z$ nuclei below the core $^{100}$Sn.
\end{abstract}

\end{frontmatter}

Many features in nuclear structure physics can be understood in term of the 
seniority coupling scheme, which was first introduced in atomic physics by 
Racah \cite{Racah43}. This scheme showed to be extremely useful for the 
classification of nuclear states in the $jj$-scheme 
\cite{Mayer50,Flo52,Talmi93}, particularly in semimagic nuclei with only 
one type of nucleons. The lowest-seniority pair (with $v=0$) has nothing 
special from a coupling point of view since the nuclear state can then
be constructed in a variety of equivalent ways through other pairs. In 
particular, the aligned like-nucleon pair coupling was proposed in Ref. 
\cite{Chen92}, which may manifest itself from the energy differences of 
mirror nuclei \cite{Len01}. The driving force behind the dominance of seniority coupling is the strong pairing interaction between like particles. 

 The neutron-proton ($np$) correlation breaks the seniority symmetry in a major way.
Correspondingly, the wave function is a mixture of many components with different seniority quantum numbers. It is not clear yet how this kind of states can be classified in the $jj$-scheme. The stretch scheme, which corresponds to the maximally aligned intrinsic angular momentum, was proposed in the 1960s to describe the rotational-like spectra of open-shell nuclei \cite{dan66,dan67}. But now it is widely accepted that a proper description of deformation involves the mixture of different orbitals.

The low-lying yrast states in $^{92}_{46}$Pd were recently 
reported  \cite{ced11}. This is the
heaviest $N=Z$ nucleus with measured  spectrum so far. It was 
suggested that in this nucleus, as well as in neighboring nuclei like 
$^{96}$Cd, the properties of the low-lying states can be largely described 
in a single $0g_{9/2}$ shell \cite{qi11,zer11}. Furthermore, it was proposed 
that the low-lying yrast states in these $N=Z$ nuclei can be
classified by a spin-aligned $np$ pair coupling scheme \cite{ced11, qi11}. 
That is, the ground state wave functions do not consist 
mainly of pairs of neutrons ($\nu\nu$) and protons 
($\pi\pi$) coupled to zero angular momenta, but rather of isoscalar 
$np$ pairs ($\nu\pi$) coupled to the maximum angular momentum 
$J$, which in the shell $0g_{9/2}$ is $J=9$   \cite{ced11, qi11}. 
A detailed shell-model analysis of the spin-aligned $np$ pair coupling was 
performed in Ref. \cite{qi11}. This shell model calculation was done in
the standard fashion of using as representation the tensorial product of
neutron times proton degrees of freedom, thus easily takes into account the
Pauli principle. The drawback with this representation is that it
is not straightforward to realize that the states are mainly determined by
$np$ pair degrees of freedom. This can only be done by projecting 
the shell model wave function into the particular $np$ component one wishes
by using two-particle coefficients of fractional 
parentage. This calculation becomes rather involved for systems with more 
than three pairs. Moreover, it does not allow one to study simultaneously the 
competing effects of different $np$ pairs, which requires a large amount of
independent projections. A similar calculation was done in Ref. \cite{zer11} 
for the two-pair case, just confirming the coupling scheme mentioned above. For 
systems with three and four pairs, the interacting boson model was applied 
through the boson mapping of the aligned $np$ pair \cite{zer11}.

In this Letter we will show that a suitable representation to study 
simultaneously different partitions of a system
consisting of many $np$ excitations is the
multistep shell model method (MSM) \cite{lio81}.  In this method one solves
the shell-model equation in several steps. In the first step one constructs
the two-particle states. In the second step one proceed by solving the
three- or four-particle states in terms of the two-particle states calculated
in the first step. In
our case we will solve the two-neutron plus two-proton system within a non-orthogonal overcomplete basis in terms of the
$(\nu\pi)\otimes(\nu\pi)$ excitations at
the same time as the $(\nu\nu)\otimes(\pi\pi)$ ones. With the four-particle
system thus evaluated, we will  proceed to evaluate the
six-particle system in terms of the coupling of the four-particle and two-particle states. For the eight-particle system one can choose 
the MSM basis such that it consists of the products of
the four-particle states in the form $(\nu\pi)\otimes(\nu\pi)$. Systems with more pairs can be described in the same fashion in 
successive steps. 

The MSM treatment of four and six identical particles was performed in Ref.
\cite{lio81}. Although the formalism to be used here is similar, the present calculation is even more challenging due to
the presence of both neutron and proton degrees of freedom. In the case of 
identical particles the MSM basis is overcomplete mainly because it
violate the Pauli principle. In our case the overcompleteness of the basis 
is even more severe since our basis elements may count twice the same states 
besides violations of the Pauli principle. For instance, in the two-pair case 
the basis elements $(\nu\nu)\otimes(\pi\pi)$ and $(\nu\pi)\otimes(\nu\pi)$ may 
be proportional to each other (see, also, Refs. \cite{qi11,zer11}). The 
overcounting thus occurring is a result of describing the $np$ and 
like-particle excitations at the same time. In the MSM this complication is 
overcome by evaluating the overlap matrix from which an orthonormal set of 
states can be constructed. This can be a very time consuming 
procedure for systems with more than two pairs. 

We will use the Greek letter $\gamma_n$ to label the $n$-particle 
$np$ states. Since we
will only consider cases with equal number of neutrons and protons outside a
closed shell, $n$ will be an even number. A $m$-proton ($m$-neutron) state will be labelled by $\alpha_m$ ($\beta_m$). 
Therefore the $np$ states will be
$|\gamma_2\rangle=P^+(\gamma_2)|0\rangle$ where the $np$ creation operator is
$P^+(\gamma_2)= \sum_{i,p}X(ip;\gamma_2)c^+_ic^+_p$ and $c^+_i$ ($c^+_p$)
is the neutron (proton) single-particle creation operator. In the same fashion the
two-proton (two-neutron) creation operator will be denoted as $P^+(\alpha_2)$
($P^+(\beta_2)$) (c.f., Eq. (9) of Ref. \cite{lio81}). The four-particle
state, $|\gamma_4\rangle=P^+(\gamma_4)\vert 0\rangle$, is
\bea
\label{eq:wf4p}
P^+(\gamma_4)&=&\sum_{\alpha_2,\beta_2}X(\alpha_2\beta_2;\gamma_4)
P^+(\alpha_2) P^+(\beta_2)
\nn &+&
\sum_{\gamma_2\leq \gamma_2'}X(\gamma_2\gamma_2';\gamma_4)
P^+(\gamma_2)P^+(\gamma_2'),
\eea
where all possible like-particle and $np$ pairs are taken into account.
Since the number of
MSM basis vectors is larger than the dimension of the shell model space, the wave function amplitudes $X$ are not well defined in our case
and, therefore,
they are not meaningful physically. The meaningful quantities
are the projections of the basis
vectors upon the physical vector, which we denote as
\bea
\label{eq:proj}
F(\alpha_2\beta_2;\gamma_4)&=&\langle\gamma_4|P^+(\alpha_2) P^+(\beta_2)|0\rangle,
\nn 
F(\gamma_2\gamma_2';\gamma_4)&=&\langle\gamma_4|P^+(\gamma_2)P^+(\gamma_2')|0\rangle.
\eea
The orthonormality condition now reads
\bea
\label{eq:ort4}
\delta_{\gamma_4\gamma_4'}&=&
\sum_{\alpha_2,\beta_2}X(\alpha_2\beta_2;\gamma_4)
F(\alpha_2\beta_2;\gamma_4') 
\nn &+&
\sum_{\gamma_2\leq \gamma_2'}X(\gamma_2\gamma_2';\gamma_4)
F(\gamma_2\gamma_2';\gamma_4).
\eea

The norm of the MSM basis
$|\gamma_2\gamma_2'\rangle$ $=$ $P^+(\gamma_2)P^+(\gamma_2')|0\rangle$, i.e.,
$N(\gamma_2\gamma_2';\gamma_4)=
\sqrt{\langle\gamma_2\gamma_2'|\gamma_2\gamma_2'\rangle}$,
may not be unity. Therefore
the interesting quantity is not the projection $F$ but rather the cosine
of the angle between the basis vector and the physical vector, i.e.,  $\cos(\phi)=x$ and 
\be
\label{cos4}
x(\gamma_2\gamma_2';\gamma_4)=
F(\gamma_2\gamma_2';\gamma_4)/N(\gamma_2\gamma_2';\gamma_4).
\ee 
If we would have taken as basis elements the complete set of
orthonormal states
$\{P^+(\alpha_2) P^+(\beta_2)|0\rangle\}$ (which is the standard shell model
basis as used in Ref. \cite{qi11}) then the second term in Eq. (\ref{eq:ort4}) would not have appeared
and one would have obtained
$X(\alpha_2\beta_2;\gamma_4)=x^*(\alpha_2\beta_2;\gamma_4)$, as expected
in an orthonormal basis.
One thus sees that the advantage of the MSM basis is that one can extract
the physical structure of the calculated states just by examining the
quantity $x$.

For the six-particle case we will use the MSM partition of two- times
four-particles, as it was done in Ref. \cite{lio81} for systems with six like 
particles. Thus the corresponding 
wave function will be $|\gamma_6\rangle=P^+(\gamma_6)|0\rangle$, where
\be
\label{eq:wf6p}
P^+(\gamma_6)=
\sum_{\gamma_2,\gamma_4}X(\gamma_2\gamma_4;\gamma_6)
P^+(\gamma_2)P^+(\gamma_4).
\ee
As before, we will evaluate the projection of the basis vectors upon the physical
vectors, i.e., $F(\gamma_2\gamma_4;\gamma_6)$,
and the corresponding cosine function $x$.
In this six-particle case one can also view the MSM basis elements as the direct
tensorial product of three pairs. This is a unique feature of the MSM. The projection of such a
MSM basis upon the physical vector is, \bea \label{eq:proj6}
F(\gamma_2\gamma_2'\gamma_2'';\gamma_6)&=& \langle\gamma_6|P^+(\gamma_2)
P^+(\gamma_2')P^+(\gamma_2'')|0\rangle \nn &=& \sum_{\gamma_4}
F(\gamma_2\gamma_4;\gamma_6)F(\gamma_2'\gamma_2'';\gamma_4), \eea 
from which one can evaluate the norm as $
N^2(\gamma_2\gamma_2'\gamma_2'';\lambda)= \sum_{\gamma_6(\lambda)}
|F(\gamma_2\gamma_2'\gamma_2'';\gamma_6)|^2 $, where $\lambda$ is the total
angular momentum of the state and the sum runs over all physical states
$\gamma_6$ with angular momentum $\lambda$. The cosine of the angle between
a MSM basis and the physical state is \be \label{cos6}
x(\gamma_2\gamma_2'\gamma_2'';\gamma_6)=
F(\gamma_2\gamma_2'\gamma_2'';\gamma_6)/N(\gamma_2\gamma_2'\gamma_2'';\lambda).
\ee 

We will describe the eight-particle states as
$|\gamma_8\rangle$ = $P^+(\gamma_8)|0\rangle$, where $P^+(\gamma_8)$ =
$\sum_{\gamma_4\leq \gamma_4'}X(\gamma_4 \gamma_4';\gamma_8)
P^+(\gamma_4)P^+(\gamma_4')$. Proceeding as above we will also evaluate the
cosine of the angle between $|\gamma_8\rangle$ and all the possible four-pair
states that can be formed.

It is important to point out that for any MSM basis
element $|b_n\rangle$ corresponding to the $n$-particle system it is
$\sum_{\gamma_n}x^2(b_n,\gamma_n)$=1. This is because the vectors
$|\gamma_n\rangle$, which are eigenvectors of the $n$-particle Shell Model
Hamiltonian, form an orthonormal (complete) set. Therefore the cosine
$x(b_n,\gamma_n)$ is the probability of the state $|\gamma_n\rangle$
occupying the basis state $|b_n\rangle$.  

We will apply the method to study the spin-aligned $np$ pair coupling scheme 
\cite{ced11,qi11}. We will restrict our calculations to the single $0g_{9/2}$ 
shell with the interaction matrix elements taken from Ref.~\cite{qi11}. But it 
should be emphasized that the formalism proposed in the present work can be 
naturally generalized to systems with many shells. 

In the cases of $^{96}$Cd and $^{92}$Pd the low-lying spectra
are determined by the isoscalar and strongly attractive matrix element
$\langle(g_{9/2})^2;9|V|(g_{9/2})^2;9\rangle$ \cite{qi11}, which 
corresponds to the maximally aligned $np$ pair configuration.
The extend to which this determines the spectrum can be deemed by the
evolution of the calculated levels as a
function of the variations of that matrix element. Calling
$V_9(\delta)= V_9(0)(1+\delta)$  it is found that
as $V_9(\delta) \rightarrow 0$ the spectrum tends to have 
a seniority-like form. At the other extreme, approaching $\delta $=1, a tendency 
towards a vibrational-like
spectrum seems to take place \cite{qi11}.
This is not surprising since, as also seen below, all low-lying yrast states in both nuclei
are isoscalar $np$ pair excitations \cite{qi11}.

\begin{figure}
\begin{center}
\begin{tikzpicture}[scale=1.0,very thick]
\begin{axis}[
xtick={-1.0,-0.5,0.0,0.5,1.0},
xmin=-1.2,xmax=1.2,ymin=0.4,ymax=1.0, minor y tick num = 1, ytickmax=0.99,minor x tick num = 1,
xlabel={\large $\delta$},
ylabel={\large$x^2$},
height=5cm,width=8cm,name=axisSnBE2,
]
\draw node at  (axis cs:-1.0,0.92){\large $2^+$};
\draw node at  (axis cs:-1.1,0.75){ 1};
\draw node at  (axis cs:-1.1,0.5){ 2};
\addplot [color=red, solid,thick,  mark=none]
coordinates {
(1,    0.99003)
(0.9,  0.99003)
(0.8,  0.98804)
(0.7,  0.98804)
(0.6,  0.98605)
(0.5,  0.98406)
(0.4,  0.98208)
(0.3,  0.9801)
(0.2,  0.97812)
(0.1,  0.97614)
(0,    0.9722)
(-0.1, 0.96826)
(-0.2, 0.96236)
(-0.3, 0.95453)
(-0.4, 0.94478)
(-0.5, 0.93316)
(-0.6, 0.91776) 
(-0.7, 0.89492)
(-0.8, 0.86304)
(-0.9, 0.81903)
(-1.0, 0.75516) 
};
\addplot [color=blue, solid,thick,  mark=none]
coordinates {
(1,    0.3969)
(0.9,  0.39816)
(0.8,  0.40069)
(0.7,  0.40196)
(0.6,  0.4045)
(0.5,  0.40704)
(0.4,  0.4096)
(0.3,  0.41216)
(0.2,  0.41602)
(0.1,  0.4199)
(0,    0.4238)
(-0.1, 0.42772)
(-0.2, 0.43296)
(-0.3, 0.43957)
(-0.4, 0.44622)
(-0.5, 0.45428)
(-0.6, 0.4624) 
(-0.7, 0.47197)
(-0.8, 0.48164)
(-0.9, 0.4914)
(-1.0, 0.49844) 
};
\end{axis}

\begin{axis}[
xtick={-1.0,-0.5,0.0,0.5,1.0},
ytick={0.4,0.6,0.8,1.0},
xticklabels={},
xmin=-1.2,xmax=1.2,ymin=0.4,ymax=1.0,ytickmax=1.0, minor y tick num = 1,minor x tick num = 1,
ylabel={\large $x^2$},
height=5cm,width=8cm,name=axisTeBE2,
at={($(axisSnBE2.north)-(0,0cm)$)},anchor=south
]
\draw node at  (axis cs:-1.0,0.92){\large $0^+$};
\draw node at  (axis cs:1.1,0.96){ 1};
\draw node at  (axis cs:1.1,0.53){ 2};
\draw node at  (axis cs:1.1,0.48){3};
\draw node at  (axis cs:1.1,0.43){ 4};
\addplot [color=red, solid,thick,  mark=none]
coordinates {
(1,     0.97417)
(0.9,   0.9722)
(0.8,   0.96826)
(0.7,   0.96629)
(0.6,   0.96236)
(0.5,   0.95844)
(0.4,   0.95258)
(0.3,   0.94673)
(0.2,   0.93896)
(0.1,   0.93122)
(0,     0.9216)
(-0.1,  0.90821)
(-0.2,  0.89302)
(-0.3,  0.87423)
(-0.4,  0.85008)
(-0.5,0.82084)
(-0.6,0.785)
(-0.7,0.7396)
(-0.8,0.68724)
(-0.9,0.6241)
(-1,0.55652)
};
\addplot [color=blue, solid,thick,  mark=none]
coordinates {
(1,    0.49562)
(0.9,  0.49985)
(0.8,  0.50552)
(0.7,  0.51123)
(0.6,  0.5184)
(0.5,  0.52563)
(0.4,  0.53436)
(0.3,  0.54317)
(0.2,  0.55205)
(0.1,  0.564)
(0,    0.57608)
(-0.1, 0.58829)
(-0.2, 0.60373)
(-0.3, 0.61937)
(-0.4, 0.6368)
(-0.5, 0.6561)
(-0.6, 0.67568) 
(-0.7, 0.69389)
(-0.8, 0.71234)
(-0.9, 0.7242)
(-1.0, 0.73274) 
};
\addplot [color=green, solid,thick,  mark=none]
coordinates {
(1,    0.45428)
(0.9,  0.45024)
(0.8,  0.44489)
(0.7,  0.43957)
(0.6,  0.43428)
(0.5,  0.42902)
(0.4,  0.4212)
(0.3,  0.41345)
(0.2,  0.40577)
(0.1,  0.39564)
(0,    0.3844)
(-0.1, 0.37088)
(-0.2, 0.35522)
(-0.3, 0.33872)
(-0.4, 0.3181)
(-0.5, 0.29485)
(-0.6, 0.26832) 
(-0.7, 0.23814)
(-0.8, 0.20521)
(-0.9, 0.16974)
(-1.0, 0.13469) 
};
\addplot [color=black, solid,thick,  mark=none]
coordinates {
(1,    0.43428)
(0.9,  0.42772)
(0.8,  0.4212)
(0.7,  0.41345)
(0.6,  0.40577)
(0.5,  0.39564)
(0.4,  0.38564)
(0.3,  0.37454)
(0.2,  0.3612)
(0.1,  0.34692)
(0,    0.33063)
(-0.1, 0.31136)
(-0.2, 0.29052)
(-0.3, 0.26729)
(-0.4, 0.2401)
(-0.5, 0.21068)
(-0.6, 0.17808) 
(-0.7, 0.14288)
(-0.8, 0.10758)
(-0.9, 0.07344)
(-1.0, 0.04326) 
};
\end{axis}
\end{tikzpicture}
\end{center}
\caption{(Color online)
Upper: Coefficients $x^2$ corresponding to the wave function of the ground state of $^{96}$Cd as a function
of the controlling parameter $\delta$ (see text).
The labels indicate 1:
$|((\nu\pi)_9)^2;0\rangle$; 2: $|(\nu\nu)_0(\pi\pi)_0;0\rangle$; 3: $|((\nu\pi)_1)^2;0\rangle$;
4: $|(\nu\nu)_2(\pi\pi)_2;0\rangle$. Lower: Same as the upper panel but for the first $2^+$ state. 
The labels correspond to 1: $|((\nu\pi)_9)^2;2\rangle$; 2: $|(\nu\pi)_0(\nu\pi)_2;2\rangle$.
Only configurations with $x^2>0.4$ are shown for simplicity.
}
\label{str0}
\end{figure}
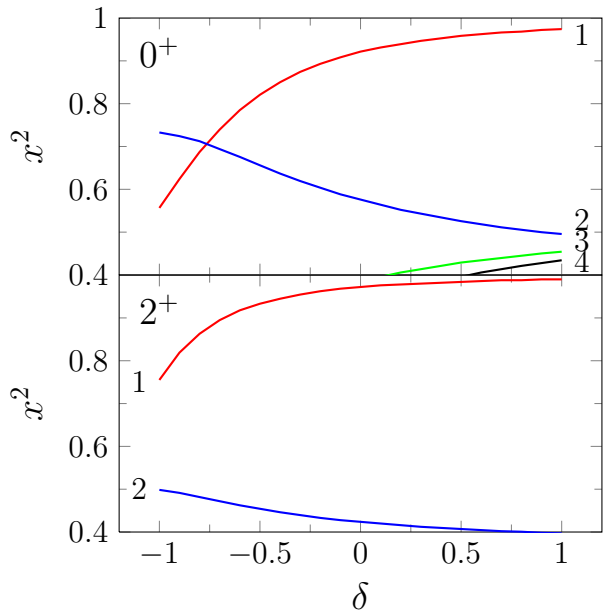

The MSM provides in a straightforward fashion the structure of the
states in terms of 
all possible configurations. Thus, in Fig.~\ref{str0} we show the
main values of the probabilities $x^2$ (Eq. (\ref{cos4})),  as a function of the controlling parameter $\delta$,  
for the ground state and first $2^+$ state of $^{96}$Cd. The striking feature in this
figure is that the spectrum for $\delta $=0 is dominated by the isoscalar
configuration $((\nu\pi)_9)^2$.  Moreover, one sees that as $V_9(\delta)$
becomes more attractive ($\delta \rightarrow 1$)
the pairing state becomes less and
less relevant while the importance of other isoscalar components increases. 
This feature is even more remarkable for the state $2^+_1$, where the spin-aligned $np$ pair coupling $((\nu\pi)_9)^2$ dominates the wave function for all values of $\delta$ shown in the figure.

One may wonder whether the pairing mode $|(\nu\nu)_0(\pi\pi)_0;0\rangle$,
which does not dominate the ground state in $^{96}$Cd, 
would be located as a state $0^+$ higher up in the
spectrum. However, this is not the case, as can be seen from Table \ref{cd96} where we listed the main components of the lowest-lying states in $^{96}$Cd for different total angular momenta. Rather it is 
distributed throughout the spectrum. In contrast, the isoscalar aligned mode 
$|((\nu\pi)_9)^2;I\rangle$ is mainly concentrated in the yrast states. 
It may seem weird that these two modes produce different results since
they may be related to each other just by an exchange of neutrons and
protons. But they are not the same, as shown by the angle $\phi$ 
between them which gives $\cos (\phi)$=0.62, i.e., $\phi = 52^{\circ}$. It is also
interesting to notice that the norm of the aligned state is
$\langle ((\nu\pi)_9)^2;0|((\nu\pi)_9)^2;0\rangle=2.00001$, which shows that the
influence of the Pauli principle upon $|((\nu\pi)_9)^2;0\rangle$ is negligible and,
therefore, it represents virtually a bosonic mode (see also Ref. \cite{zer11}). 

\begin{table*}
\begin{center}
\caption{Leading configurations in the first five states of $^{96}$Cd for a given total angular momentum $I$.}
\label{cd96}
\vskip 2mm
\scalebox{0.7}{ 
\begin{tabular}{cccccccccccc}
\hline
&\multicolumn{2}{c}{$I=0$} &&\multicolumn{2}{c}{$I=2$} &&\multicolumn{2}{c}{$I=4$} &&\multicolumn{2}{c}{$I=6$} \cr
\cline{2-3}\cline{5-6}\cline{8-9}\cline{11-12}
n& Configuration   & $|x|$ &&Configuration& $|x|$&&Configuration& $|x|$&&Configuration& $|x|$\cr
\hline
1&$(\nu\pi)_9\otimes(\nu\pi)_9$& 0.96&&$(\nu\pi)_9\otimes(\nu\pi)_9$& 0.99  &&$(\nu\pi)_9\otimes(\nu\pi)_9$& 0.97&&$(\nu\pi)_9\otimes(\nu\pi)_9$& 0.84\cr
2&$(\nu\pi)_5\otimes(\nu\pi)_5$& 0.88 &&$(\nu\pi)_8\otimes(\nu\pi)_9$& 0.92
&&$(\nu\pi)_8\otimes(\nu\pi)_9$& 0.93&&$(\nu\pi)_7\otimes(\nu\pi)_9$& 0.60\cr
3&$(\nu\pi)_0\otimes(\nu\pi)_0$& 0.77&& $(\nu\pi)_7\otimes(\nu\pi)_9$& 0.94
&&$(\nu\pi)_7\otimes(\nu\pi)_9$& 0.90&&$(\nu\nu)_{0(6)}\otimes(\pi\pi)_{6(0)}$& 0.94\cr
4&$(\nu\pi)_3\otimes(\nu\pi)_3$& 0.91 && $(\nu\pi)_7\otimes(\nu\pi)_8$& 0.82
&&$(\nu\pi)_5\otimes(\nu\pi)_9$& 0.90&&$(\nu\nu)_{2(6)}\otimes(\pi\pi)_{6(2)}$& 0.65\cr
5&$(\nu\pi)_4\otimes(\nu\pi)_4$& 0.80 && $(\nu\pi)_5\otimes(\nu\pi)_7$& 0.64
&&$(\nu\pi)_6\otimes(\nu\pi)_9$& 0.92&&$(\nu\nu)_{2(6)}\otimes(\pi\pi)_{6(2)}$& 0.84\cr
\hline
\end{tabular}
}
\end{center}
\end{table*}

\begin{figure}
\begin{center}
\includegraphics[width=13cm]{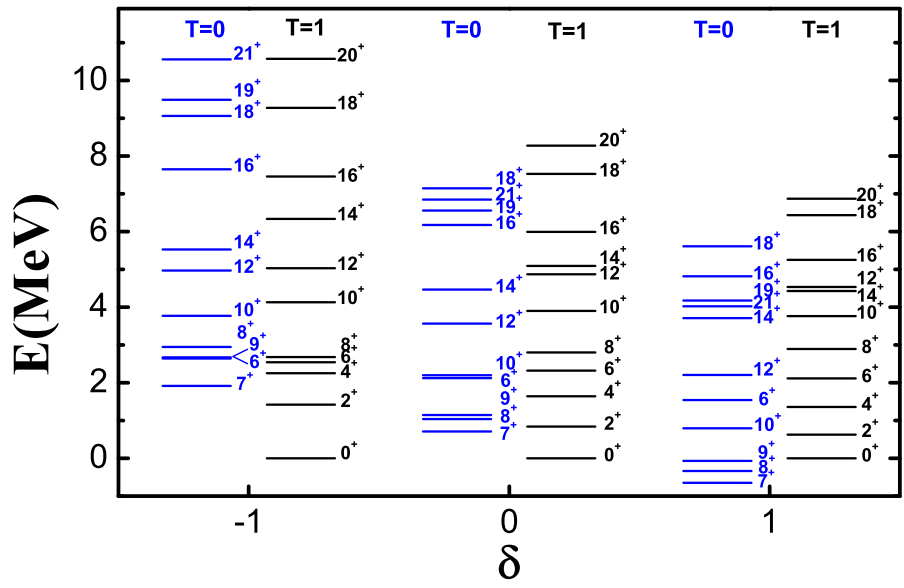}\\
\end{center}
\caption{(Color online)
Shell model spectra of $^{94}$Ag calculated in the ${0g_{9/2}}$ shell as a function of 
the controlling parameter $\delta$.
}
\label{ene94}
\end{figure}

The calculated spectrum of the odd-odd nucleus $^{94}_{47}$Ag is shown in Fig. 
\ref{ene94}, where we have grouped the levels according to their isospin.
There are only two states which have been measured in this case
\cite{ag24}, namely the $0^+$ ground state and a state
$21^+$ at 6.67 MeV. These states are in reasonable
agreement with experiment and coincide with previous 
shell model calculations \cite{com02}. The strong influence of the aligned isoscalar matrix element $V_9$ upon the $T=0$ 
states can be inferred from the figure, where it is seen that 
the energies of 
$T=0$ states are much more sensitive to the controlling parameter $\delta$ than those with $T=1$. 

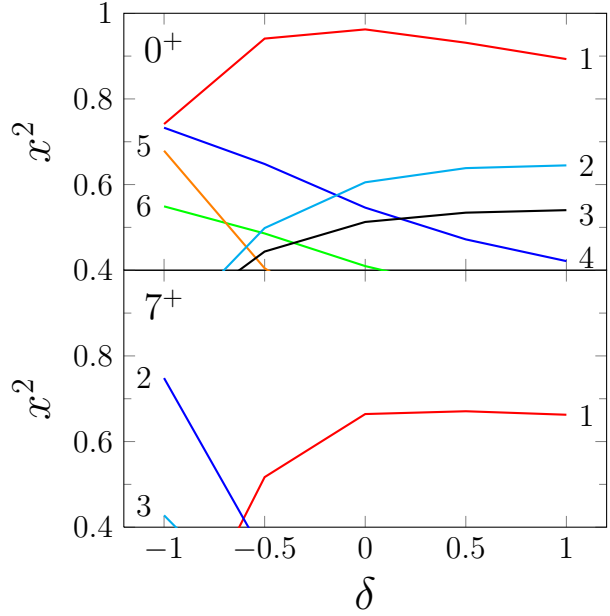
\begin{figure}
\begin{center}
\begin{tikzpicture}[scale=1.0]
\begin{axis}[
xtick={-1.0,-0.5,0.0,0.5,1.0},
xmin=-1.2,xmax=1.2,ymin=0.4,ymax=1.0, minor y tick num = 1, ytickmax=0.99,
xlabel={\Large $\delta$},
ylabel={\Large$x^2$},
height=5cm,width=8cm,name=axisSnBE2,
]
\draw node at  (axis cs:-1.0,0.92){\large $7^+$};
\draw node at  (axis cs:1.1,0.66){ 1};
\draw node at  (axis cs:-1.1,0.75){ 2};
\draw node at  (axis cs:-1.1,0.45){ 3};
\addplot [color=red, solid,thick,  mark=none]
coordinates {
(-1,    0.054756 )
(-0.5,  0.516961 )
(0,     0.664225 )
(0.5,   0.670761 )
(1,     0.662596 )
};\addplot [color=blue, solid,thick,  mark=none]
coordinates {
(-1,    0.748225 )
(-0.5,  0.332929 )
(0,     0.137641 )
(0.5,   0.103041 )
(1,     0.090601 )
};\addplot [color=cyan, solid,thick,  mark=none]
coordinates {
(-1,    0.427716)
(-0.5,  0.190096)
(0,     0.078961)
(0.5,   0.059049)
(1,     0.051529)
};

\end{axis}

\begin{axis}[
xtick={-1.0,-0.5,0.0,0.5,1.0},
ytick={0.4,0.6,0.8,1.0},
xticklabels={},
xmin=-1.2,xmax=1.2,ymin=0.4,ymax=1.0,ytickmax=1.0, minor y tick num = 1,
ylabel={\Large $x^2$},
height=5cm,width=8cm,name=axisTeBE2,
at={($(axisSnBE2.north)-(0,0cm)$)},anchor=south
]
\draw node at  (axis cs:-1.0,0.92){\large $0^+$};
\draw node at  (axis cs:1.1,0.893){ 1};
\draw node at  (axis cs:1.1,0.645){ 2};
\draw node at  (axis cs:1.1,0.53){3};
\draw node at  (axis cs:1.1,0.43){ 4};
\draw node at  (axis cs:-1.1,0.69){5};
\draw node at  (axis cs:-1.1,0.55){ 6};
\addplot [color=red, solid,thick,  mark=none]
coordinates {
(-1,    0.741321 )
(-0.5,  0.9409 )
(0,     0.962361 )
(0.5,   0.931225 )
(1,     0.893025 )
};
\addplot [color=blue, solid,thick,  mark=none]
coordinates {
(-1,    0.732736 )
(-0.5,  0.648025 )
(0,     0.546121)
(0.5,   0.471969 )
(1,     0.421201 )
};
\addplot [color=green, solid,thick,  mark=none]
coordinates {
(-1,   0.549081)
(-0.5, 0.485809)
(0,    0.4096)
(0.5,  0.354025)
(1,    0.315844)
};
\addplot [color=orange, solid,thick,  mark=none]
coordinates {
(-1,    0.678976 )
(-0.5,  0.404496 )
(0,     0.258064 )
(0.5,   0.183184 )
(1,     0.142129 )
};
\addplot [color=cyan, solid,thick,  mark=none]
coordinates {
(-1,    0.255025)
(-0.5,  0.498436)
(0,     0.605284)
(0.5,   0.638401)
(1,     0.644809)
};
\addplot [color=black, solid,thick,  mark=none]
coordinates {
(-1,   0.2704 )
(-0.5, 0.443556)
(0,    0.512656 )
(0.5,  0.534361 )
(1,    0.540225 )
};
\end{axis}
\end{tikzpicture}
\end{center}
\caption{(Color online)
Upper: Coefficients $x^2$ corresponding to the wave function of $^{94}$Ag ($0^+_1$) as a function
of $\delta$.
The labels indicate  
1: $|(\nu\pi)_0(\nu\pi)_9(\nu\pi)_9);0\rangle$; 
2: $|(\nu\pi)_0(\nu\nu)_2(\pi\pi)_2);0\rangle$;
3: $|(\nu\pi)_1(\nu\nu)_8(\pi\pi)_8);0\rangle$;
4: $|(\nu\pi)_0(\nu\pi)_0(\nu\pi)_0);0\rangle$;
5: $|(\nu\pi)_0(\nu\pi)_5(\nu\pi)_5);0\rangle$;
6: $|(\nu\pi)_0(\nu\nu)_0(\pi\pi)_0);0\rangle$. Lower: Same as the upper panel but for the $7^+_1$ state. The labels indicate
1: $|(\nu\pi)_0(\nu\pi)_9(\nu\pi)_9);7\rangle$; 2: $|(\nu\pi)_7(\nu\nu)_0(\pi\pi)_0);0\rangle$;
3: $|(\nu\pi)_7(\nu\pi)_0(\nu\pi)_0);7\rangle$.
}
\label{str06}
\end{figure}

The  $T=0$ states in $^{94}$Ag are specially interesting because in
this case the MSM basis vector consisting of the three $9^+$ isoscalar aligned 
states is not hindered by any symmetry (recall that only  states with total isospin $T=0$ can be coupled from the isoscalar $np$ pairs). 
Indeed we found that most of the yrast levels in Fig. 
\ref{ene94}, except the $16^+_1$ and $18^+_1$ states,  are mainly built by the isoscalar aligned $np$ pairs. As an example, in the lower panel of Fig.  \ref{str06} we present the  the  probabilities
$x^2$ for main components of the $7^+_1$ state of $^{94}$Ag, which is calculated to be the lowest $T=0$ state, as a function of the controlling parameter $\delta$.

It is seen from Fig. 
\ref{ene94} that at $\delta=0$
the ground state of $^{94}$Ag carries $T=1$. It may thus seem
that in odd-odd system the isovector pairing mode
retakes its predominance. We found that this is not the case by analyzing the
ground state wave function in terms of the tensor product of three pairs, as can be seen from
the upper panel of Fig. \ref{str06}.
The most important configuration consists of the spin-aligned
$np$ pair state $|(\nu\pi)_9(\nu\pi)_9);0\rangle$ that was dominant in $^{96}$Cd and the
other possible configuration, i.e., the isovector state $|(\nu\pi)_0\rangle$. 
In fact
the low-lying $T=1$ states here have the same origin as the corresponding $T=1$
states in $^{94}$Cd. 
One sees that in this odd-odd six-particle case the probabilities of
the ground state occupying different basis states are larger than in 
$^{96}$Cd, Fig. \ref{str0}. This is not surprising since  for systems with more than two pairs there are many nearly equivalent combinations that can be built in the same fashion.

In the analysis of eight-particle systems like $^{92}$Pd we choose as MSM basis the partition 
$|\gamma_4\gamma_4';\gamma_8\rangle$. 
In this case the MSM basis is highly overcomplete. For
instance there are 36 shel-model $0^+$ states while the corresponding MSM dimension is
915.
Within this basis we calculated in Table \ref{fbos0} the quantities $x$, i.e., the cosines 
of the angles between the vectors 
$|\gamma_8\rangle$ and all the possible vectors  that can be formed by the coupling of
four pairs. Since many
combinations are similar to each other there is not a value of $x$ which is significantly
larger than the others. 
But one finds, again, that
for the ground state of $^{92}$Pd the most important MSM configuration
is the one corresponding to
the four $9^+$ aligned pairs. The second one is a combination of two
aligned $9^+$ states and the normal pairing states. This is expected since in the two-pair case of $^{96}$Cd the second largest component is the normal pairing term. An important feature in this case is that for the pairing state it is
$x^2(\alpha_2=0^+,\beta_2=0^+\alpha_2'=0^+\beta_2'=0^+;\gamma_8=0^+_1)$
= 0.46. This is a relatively small number. Indeed it occupies the 
10th place in order of importance. This reflects, once again, the
dominance of the aligned configuration in this nuclear region. 

\begin{table}
\begin{center}
\caption{Configurations with the largest probabilities for the state $^{92}$Pd($0^+_1$)
corresponding to the tensorial products of different two-particle
states (upper) and four-particle states (lower).}
\label{fbos0}
\vskip2mm
\begin{tabular}{c|c}
\hline
Configuration   & $x^2$ \cr
\hline
$|\gamma_2=9^+\gamma_2'=9^+\gamma_2''=9^+\gamma_2'''=9^+\rangle$& 0.85  \cr
$|\gamma_2=9^+\gamma_2'=9^+\alpha_2=0^+\beta_2=0^+\rangle$& 0.76  \cr
$|\gamma_2=8^+\gamma_2'=1^+\alpha_2=0^+\beta_2=8^+\rangle$& 0.56  \cr
$|\gamma_2=8^+\gamma_2'=1^+\alpha_2=8^+\beta_2=0^+\rangle$& 0.56  \cr
$|\gamma_2=1^+\gamma_2'=1^+\alpha_2=0^+\beta_2=0^+\rangle$& 0.52  \cr
\hline
\end{tabular}
\end{center}
\end{table}

Summarizing, we have in this paper extended the MSM method proposed in 
Ref. \cite{lio81} to incorporate both neutron and proton degrees of freedom 
and applied it to study the recently proposed spin-aligned $np$ pair coupling 
scheme. 
We have applied the method to analyze four-, six- and eight-hole states in
$N=Z$ nuclei below the core $^{100}$Sn. The calculations were performed
within the restricted $0g_{9/2}$ shell for simplicity. But this work opens the 
way for even more challenging calculations, involving many particles and/or 
many shells. It would also allow one to truncate the shell model basis in 
terms of spin-aligned $np$ pairs or other coupling schemes.

This work was supported by the Swedish Research Council (VR) under grant Nos. 623-2009-7340 and 2010-4723. Z.X. is supported in part by the China Scholarship Council under
grant No. 2008601032.

\end{document}